\documentclass[11pt,leqno,textwidth=10cm]{amsart}
\usepackage{amsmath,amsthm,amssymb,verbatim,amsfonts}

\usepackage{mathrsfs}
\usepackage{mathtools}
\usepackage{abstract}

\usepackage{color}
\usepackage{marginnote}
\usepackage{tensor}

\usepackage{enumerate}

\usepackage{soul}


\usepackage{etoolbox}
\makeatletter
\patchcmd{\@maketitle}{\newpage}{}{}{} 
\makeatother

\newtheoremstyle{fancy}{}{}{\itshape}{}{\textsc\bgroup}{.\egroup}{ }{}
\newtheoremstyle{fancy2}{}{}{\rm}{}{\textsc\bgroup}{.\egroup}{ }{}
\theoremstyle{fancy}
\newcounter{intro}

\numberwithin{equation}{section}    

\newtheorem{thm}[equation]{Theorem}

\newtheorem{named}[equation]{\name}
\newcommand{\name}{Proof of}

\theoremstyle{fancy2}

\newcounter{axa}

\newtheorem{rem}[equation]{Remark}

\newcommand{\cref}[1]{Corollary~\ref{#1}}

\newcommand{\tr}{\operatorname{tr}}

\newcounter{subequation} 
 \newlength\mtabskip\mtabskip=-1.25cm
   
	 \def\mtabLong{long} 
 

\newcommand{\eq}[1]{\begin{equation}#1\end{equation}}

\newcommand{\alg}[1]{\begin{aligned}#1\end{aligned}}

\newcommand{\bc}{\begin{cases}}
\newcommand{\ec}{\end{cases}}

\newcommand{\p}[1]{\partial_{#1}}

\newcommand{\si}{\sigma}

\newcommand{\Si}{\Sigma}

\newcommand{\mbf}[1]{\mathbf{#1}}

\def\tr{\ensuremath {\mbox{tr}}}

\newcounter{mnotecount}[section]
\newcommand{\nn}{\nonumber}
\newcommand{\red}[1]{{#1}}

\newcommand{\mnote}[1]
{\protect{\stepcounter{mnotecount}}$^{\mbox{\footnotesize
$
\bullet$\themnotecount}}$ \marginpar{
\raggedright\tiny\em
$\!\!\!\!\!\!\,\bullet$\themnotecount: #1} }

\begin{document}


\title[Isotropization of slowly expanding spacetimes]{Isotropization of slowly expanding spacetimes}
\author[\textsc{H.~Barzegar, D.~Fajman, G.~Hei\ss el}]{\textsc{Hamed Barzegar, David Fajman, Gernot Hei\ss el}}
\address{
Gravitational Physics\\
Faculty of Physics\\
University of Vienna\\
Boltzmanngasse 5, 1090 Vienna \\
Austria}
\email{Hamed.Barzegar@univie.ac.at, David.Fajman@univie.ac.at, Gernot.Heissel@univie.ac.at}

\thanks{The authors acknowledge the support of the Austrian Science Fund (FWF) via the project \emph{Geometric transport equations and the non-vacuum Einstein flow} (P 29900-N27).}

\date{\today}


\maketitle
\begin{abstract}
We show that the homogeneous, massless Einstein-Vlasov system with toroidal spatial topology and diagonal Bianchi type I  symmetry for initial data close to isotropic data isotropizes towards the future and in particular asymptotes to a radiative Einstein-deSitter model. We use an energy method to obtain quantitative estimates on the rate of isotropization in this class of models. 
\end{abstract}

\section{Introduction}
Determining the asymptotic behaviour for cosmological models is a fundamental objective of mathematical cosmology. A lot of effort has been made to investigate the stability of the isotropy of the universe (cf., e.g., \cite{AS91, Ba14, BK01a, BK01b, CH73, LBBS97}, and references therein). For the Einstein-Vlasov system, which models universes containing ensembles of self-gravitating collisionless particles \cite{An11, Re04, Re08}, this program is quite advanced for the class of spatially homogeneous (SH) spacetimes.

Much of the respective literature is based on an approach by Rendall~\cite{Re96} as well as Rendall and Tod~\cite{RT99} in which certain symmetries are imposed on the Vlasov matter distribution, as a consequence of which, together with spatial homogeneity, the Einstein-Vlasov system reduces to a system of autonomous (time-invariant) ODEs. Hence, dynamical systems theory can be applied to analyze these systems. The task to utilize this approach for all those types of SH cosmologies to which it is applicable has been acomplished in~\cite{CH09, CH10, CH11, HU06, He12, RU00} and recently been completed in~\cite{FH19}. To the latter source we also refer to for a recent and more detailed summary of this approach.

The dynamical systems approach is powerful in particular in that it is capable of yielding global stability results. It however has some 
detriments: Firstly, its results are not of full generality since it relies on the above mentioned symmetry assumptions on the matter distribution. Secondly, it cannot be applied to all types of SH cosmologies, since not for all these concrete symmetry assumptions are known, or can be found in principle. Finally, its applicability is limited to the spatially homogenous context since for cosmologies with less spatial symmetry the Einstein-Vlasov system obiously does not reduce to ODEs.

Hence, there has been an interest in adopting other techniques as well. Nungesser~\cite{Nu10, Nu12, Nu13} and Nungesser~et.~al.~\cite{Nu14} performed a small data future stability analysis based on bootstrapping techniques for several types of spatially homogenous Einstein-Vlasov cosmologies (cf.~also the result~\cite{LN17} on the Einstein-Boltzmann system.) While the stability results obtained by this approach concern generally the small data regime, it is fit to overcome the above detriments of the dynamical systems approach. The former thus complements the latter in the SH context, and leaves open the possibility of generalization to spacetimes with less spatial symmetry. The focus of this literature has been on the future stability in the case of massive particles. The dynamics in the massless case is of interest to be analyzed separately since it has been shown to behave substantially different from the massive case for various spatial topologies.

In the present paper we prove the isotropization of small perturbations of the Einstein-deSitter (EdS) model within the class of diagonal Bianchi type~I solutions to the massless Einstein-Vlasov system. Though the Bianchi~I Einstein-Vlasov system has already been investigated thoroughly in~\cite{Re96, HU06} our results are novel and complement the latter two in the following points: Firstly, the result of~\cite{Re96} is limited to the massive case with regards to the future asymptotics, while we treat the massless case. Secondly, we use an energy method by which we not only recover the result of~\cite{HU06} with regards to future asymptotics, but which also captures the decay rates for the perturbations away from isotropy -- the monotone function techniques used in~\cite{HU06} did not. Finally, we expect the energy method to also be applicable to the non-diagonal case, and that it is sufficiently robust to allow for a generalization to the class of inhomogeneous solutions. \\

We start out in Section~\ref{S: EdS} with some background on the radiative EdS model tailored to the present context. In Section~\ref{S: BI EV system} we lay out the setup for our analysis on the diagonal Bianchi~I Einstein-Vlasov system, and we formulate our result in~Theorem~\ref{T: theorem}. After a brief Section~\ref{S: strategy} on the strategy of our proof, we then present the latter in sections~\ref{S: linearization} and~\ref{S: proof}.

\section{The radiative Einstein-deSitter model}\label{S: EdS}

The radiative EdS model \red{(cf.~e.g.~\cite{We08})}
\eq{
((0,\infty) \times \mathbb T^3,-dt^2+t \cdot \gamma),
}
where $(\mathbb T^3,\gamma)$ is a flat torus, is a solution to the Einstein equations coupled to a radiation fluid, i.e.~a perfect fluid with pressures equal to $1/3$ times the energy density. With a scale factor $a(t)=\sqrt t$ it expands significantly slower than the related FLRW vacuum solution on hyperbolic spatial topologies (the Milne model) with $a(t)=t$ and also slower than the corresponding solution on $\mathbb T^3$ for dust (i.e., a pressure-less perfect fluid), the Einstein-deSitter model, with $a(t)=t^{2/3}$; cf.~\cite{Re08}. The Einstein-deSitter models pose interesting examples of \emph{matter-dominated} cosmological spacetimes, i.e.,~spacetimes whose asymptotic behaviour is altered by the presence of matter.\\
While for initial data close to the Milne geometry on hyperbolic spatial manifolds vacuum and non-vacuum future asymptotics are similar \cite{AF17}, on toroidal spatial topologies vacuum asymptotics deviate drastically from the matter dominated regime of the Einstein-deSitter models \cite{Re08}.
It is of essential interest to investigate the stability properties of those model solutions in order to understand whether their behaviour is representative for generic spacetimes with similar initial data. The stability properties of Einstein-deSitter models are unknown except in the homogeneous context, i.e.,~for Bianchi type I models. In that case it has been shown by Nungesser that the massive Einstein-deSitter model is a future attractor of the Einstein-Vlasov system with massive particles \cite{Nu10}. The analogous problem for the radiative Einstein-deSitter model, which concerns massless particles (or radiation) is addressed for the massless Einstein-Vlasov system in the present paper for the restricted class of diagonal Bianchi type I models. We show that initial data sufficiently close to an isotropic state for the massless Einstein-Vlasov system isotropizes towards the future and asymptotes towards a member of the family of radiative EdS models with suitable decay rates for the perturbations. 

The slower expansion rate for the radiative case makes it a priori more difficult to establish sufficiently strong decay estimates. We point out that nonlinear stability results are established for exponential scale factors \cite{Ri13} or polynomial scale factors with significantly higher exponents \cite{AF17}. Indeed, our analysis requires a more careful treatment of the evolution equation for the shear tensor and metric perturbations. We use a fine-tuned corrected energy, which controls shear tensor and metric perturbation simultaneously, to obtain the crucial decay estimates. 

Our theorem assures that the radiative Einstein-deSitter model is an attractor in the restricted class of diagonal Bianchi~I symmetric solutions to the massless Einstein-Vlasov system. Inhowfar stability holds in less restricted sets of solutions as for instance the set of surface-symmetric or $\mathbb T^2$-symmetric solutions is an open problem that can be addressed using the framework of previous works as for instance \cite{ARW04} and will be subject of future studies.


\section{The diagonal Bianchi type I Einstein--Vlasov System}\label{S: BI EV system}

In much of this section we closely follow section~2 of \cite[Sec~2]{HU06}, and at the end of this section we formulate our theorem and emphasize the relation to the result of this source. For a deeper background on the Bianchi classification and on the choice of basis we refer to~\cite{WE97}. For a thorough background on the Einstein-Vlasov system we refer to~\cite{An11, Re04, Re08}. \\

Bianchi spacetimes admit a Lie algebra of Killing vector fields $K_1$, $K_2$, and $K_3$ which are tangent to the orbits of the group which is identified with the universal covering space of Bianchi models. These orbits are called surface of homogeneity. Moreover, the Killing vector fields satisfy the commutation relation $ [K_i, K_j] = C^k_{ij} \, K_k$ where $C^k_{ij}$ are structure constants. Bianchi I models are characterized by $C^k_{ij}=0$. Choosing a unit vector field $n$ normal to the group orbits, one has natural choice for the time coordinate. One can choose a basis $\{E_i\}$ of the surfaces of homogeneity such that they commute with the Killing vector fields. In this way, one can construct the so-called left-invariant frame $\{n,E_i\}$ which is generated by the right-invariant Killing vector fields; cf.~\cite[Sec~1.5.2]{WE97}.
We now consider general Bianchi I spacetimes of the form $\overline g=-dt^2+\mbf g$, where $\mbf g=g_{ij}(t) W^i\otimes W^j$, where $W^i$ denote the dual one-forms to the left-invariant basis $E_i$. We denote by $k_{ij}$ the second fundamental form and decompose via
$k_{ij}=\si_{ij}-Hg_{ij}$, where $H=-\tfrac{1}{3}\tr_gk$ and $\si_{ij}$ is the trace-free part of $k_{ij}$. Moreover, we define the rescaled trace-free part and its square by
\eq{
\Si_{i}^j :=H^{-1}\si_i^j
\quad \text{and} \quad
F:=\Si_i^j\Si_j^i
\,,
}
respectively.
The physical interpretation of the defined quantities is as follows: $H$ is the Hubble scalar and represents a measure of the overall, isotropic, rate of spatial expansion. $\si_{ij}$ is the shear tensor and represents a measure of the anisotropic rate of spatial expansion. Respectively $\Si^j_i$ is the Hubble normalized shear tensor, and consequently $F$ represents an overall measure of anisotropy. In particular, $F=0$ marks an isotropic state.

As matter model we consider a collision-less kinetic gas of massless particles, i.e., massless Vlaosv matter. The respective energy-momentum tensor is determined by a distribution function $f$ which solves a transport equation – the Vlasov equation. $f(t,x^i,p^i)$ represents the density of particles at time $t$ and position $x^i$ with momentum $p^i$. In the case of Bianchi I the Vlasov equation reduces to
\begin{equation}\label{1VIII19.1}
 \p tf+2k_j^ip^j\p {p^i} f=0 .
\end{equation}
In particular, compatibility with spatial homogeneity forces $f$ to be independent of the spatial coordinates, i.e. $f=f(t,p^i)$, and hence \eqref{1VIII19.1} does not contain any spatial derivatives. The Vlasov equation \eqref{1VIII19.1} has the general spatially homogeneous solution in Bianchi type I symmetry of the form (cf. \cite[Sec~4]{MM90})
\begin{equation}
 f(t, p^i) = f_0(p_i)
 \,.
\end{equation}

Decomposing the energy momentum tensor into its spatial part $S_{ij}$, the energy density $\rho$ and the momentum density $j_k$ it is given by
\begin{eqnarray}
 \rho 
 &:=&
  \int_{\mathbb R^3\setminus\{0\}} f_0(p_k) |p|_g \sqrt{\det g} \, d^3p \,, 
  \\
S_{ij}
&:=& \int_{\mathbb R^3\setminus\{0\}} f_0(p_k) \frac{p_i p_j}{|p|_g} \sqrt{\det g} \, d^3p
\\
j_i 
&:=&
 \int_{\mathbb R^3\setminus\{0\}} f_0(p_k)  p_i \,  \sqrt{\det g} \, d^3p
 \,,
\end{eqnarray}
where $|p|_g := ( g_{ij}p^ip^j )^{1/2}$ and $d^3p := dp^1 dp^2 dp^3$. In Bianchi type I the momentum constraint dictates $j_k = 0$.

In the present paper we restrict ourselves to the subclass of Bianchi type I Einstein--Vlasov system which admits the reflection symmetry (or diagonality) in the following sense (cf.~\cite{Re96, HU06}): on the initial data the following conditions are imposed
\begin{subequations}
\begin{eqnarray}
 f_0(p_1, p_2, p_3)
 &=&
  f_0(p_1, -p_2, -p_3)
=
 f_0(-p_1, -p_2, p_3)
=
 f_0(-p_1, p_2, -p_3)
 \,,
 \\
 g_{ij}(t_0) &=& \text{diag}(g_{11}(t_0), g_{22}(t_0), g_{33}(t_0)) \,, 
 \\
 k_{ij}(t_0) &=& \text{diag}(k_{11}(t_0), k_{22}(t_0), k_{33}(t_0)) \,. 
\end{eqnarray}
\end{subequations}
Under these conditions $j_k = 0$ is satisfied and $S_{ij}(t_0)$ is diagonal. The evolution equations preserve the diagonality for all time, i.e., $g_{ij}$, $k_{ij}$, and $S_{ij}$ are diagonal for all time.

Next, we define the dimensionless variables following \cite{HU06}. We define (no summation on $i$, unless explicitly mentioned)
\begin{equation}
 \Omega
 :=
 \frac{8 \pi \rho}{3 H^2}
 \,,
 \quad
 x:= g^{11} + g^{22} + g^{33}
 \,,
 \quad
 s_i := \frac{g^{ii}}{x}
 \,,
 \quad
 \Si_i
 :=
 -\Si^i_i
 =
 - \frac{k^i_i}{H} - 1
 \,,
\end{equation}
with
$$
 s_1 + s_2 + s_3 = 1
 \,,
 \quad
 \Si_1 + \Si_2 + \Si_3 = 0
 \,. 
$$
Denote
\begin{equation}
 w_i := \frac{S^i_i}{\rho}
 \,,
 \quad
 w:= \frac{1}{3} \sum_i w_i
 \,.
\end{equation}
In the case at hand, i.e., in the massless case we have $w=1/3$. In terms of these variable $w_i$ can be written as (cf.~(9) in \cite{HU06})
\begin{equation}
 w_i
 =
 \frac{s_i \int f_0 p_i^2 \left( \sum_k s_k p_k^2 \right)^{-\frac{1}{2}} d^3\tilde{p}}{ \int f_0  \left( \sum_k s_k p_k^2 \right)^\frac{1}{2} d^3\tilde{p}}
 =:
 s_i \, Y_i (s_k)
 \,.
\end{equation}

Here, $d^3\tilde{p}=dp_1 dp_2 dp_3$. The system of constraint and evolutions equations finally takes the following form using $\partial_\tau = H^{-1} \partial_t$ (cf. \cite{HU06})
\begin{subequations}
\begin{eqnarray}
\Omega
&=&
1- \frac{1}{6} F 
\,,
\\
H'
&=&
-3 H [1 - \frac{\Omega}{2}(1-w)]
\,, \label{1VIII19.2}
\\
\Si'_i
&=&
-3 \Omega
 \left[
    \frac{1}{3} \Si_i - (w_i - w)
 \right]   
 =
 - \Omega (\Si_i + 1 - 3 w_i)
\,,
\\
 s'_i
 &=&
 -2 s_i \left( \Si_i  - \sum_k s_k \Si_k \right)
 \,,
\end{eqnarray}
\end{subequations}
where 
 `` $' \equiv \partial_\tau$."

For the system introduced above we prove the following theorem.

\begin{thm}\label{T: theorem}
Consider $C^\infty$ initial data for the massless Einstein-Vlasov system with diagonal Bianchi I symmetry, $(g_0,H_0, F_0,\rho(f_0))$ at $t_0=(2H(t_0))^{-1}$ with $f_0$ sufficiently close to an isotropic distribution function. There exists an $\varepsilon>0$ such that $F_0 <\varepsilon$ 
implies the following future asymptotics for a constant $C>0$,
\eq{\alg{
F(t)&\lesssim  \varepsilon t^{-1.16/2+\varepsilon } \,. \\
2t\leq H^{-1}(t)&\leq 2t (1+C\varepsilon t^{-1/2}) \,.
}}
and 
\eq{
tg^{ij}\rightarrow g^{ij}_\infty \mbox{ as } t\rightarrow \infty \mbox{  with  } |g_{\infty}^{ij}-t_0g_0^{ij}|\lesssim \varepsilon t^{-1.16/4+\varepsilon}. 
}
In particular, the rescaled square of the shear-tensor, $F=\red{|\Si|_g^2}$ vanishes asymptotically, i.e.,~the spacetime isotropizes. Moreover, the rescaled spatial metric $t^{-1} g(t)$ converges to a limit metric $g_{\infty}$, which remains $\varepsilon$-close to the initial metric $g_0$. In result, the radiative EdS model is orbitally stable in the set of solution to the massless Einstein-Vlasov system with diagonal Bianchi I symmetry.
\end{thm}

\begin{rem}
The choice of initial time $t_0=(2H(t_0))^{-1}$ is made for technical reasons, and does not restrict the generality.
\end{rem}

\begin{rem}
The value of of the decay rate ``$1.16$" is approximated. We obtain this value numerically from a solution of an algebraic equation (cf.~\eqref{1VIII19}). We do not claim this decay rate is sharp, but it is the optimal value in the scope of the method we use.
\end{rem}

\begin{rem}
The orbital stability in the previous theorem was already proven by Heinzle and Uggla in \cite{HU06}. However, their proof did not provide decay rates for the perturbations. Moreover, the dynamical systems method utilized in \cite{HU06} does not provide a natural extension to the inhomogeneous case, while the energy methods are flexible in their application and in principle extend to less symmetric scenarios.
\end{rem}

\section{Strategy of proof}\label{S: strategy}
In the remainder of this paper we proof the foregoing theorem. The first crucial observation concerns the attractor geometry. Namely, the tuple $\vec{s}^*:=(s^*_1,s^*_2,s^*_3)\in\mathbb R^3$, which represents the rescaled 3-metric for which, given the prescribed initial particle distribution $f_0$, $w_i(\vec{s}^*)=1/3$ holds. It has been shown by Heinzle and Uggla that $\vec{s}^*$ is unique (cf.~\cite{HU06}). Hence only these values of $s_i$ can be an attractor of this system.

We then linearize around the point $(\vec s=\vec s^*, \Sigma=0)$. The higher order terms are at least second order in the perturbation and are treated as error terms, which can be absorbed in the final energy estimate. It is the linear terms that determine the decay rates for small initial data. 

In the linearization, certain factors remain that depend on the initial particle distribution. We evaluate them for the case of isotropic initial particle distribution and conclude that their actual values are $\varepsilon$-close to those values by continuity since all functionals are continuous in $f_0$.  

After fixing all values in the linearized evolution equations for $\vec s$ and $\Sigma$ we make an ansatz for the energy, which includes a diagonal term. This ansatz is formulated in terms of two variables $\alpha$ and $\beta$, for which we derive necessary conditions for positive definiteness and decay of the energy. Then we maximize the decay rate and obtain exact values of those constants that realize the optimal energy, which proves the result. 

This type of energy methods has been used for the Einstein equations in different contexts (cf.~\cite{AF17} and references therein). Here it is used in a context where the behaviour of the matter variables has a strong effect on the geometry (matter dominated regime). We expect that it can be used in different classes of Bianchi models containing Vlasov matter. The case of Bianchi type II symmetry is currently work in progress \cite{BF19}.

\section{Linearization}\label{S: linearization}

Let $\bar{s}_i := s_i - s^\ast_i$, where $s^\ast_i$ is the component of the unique vector $\vec{s}^*$. The reduced system of evolution equations, i.e., the part of it which is decoupled of~\eqref{1VIII19.2}, then reads
\begin{eqnarray*}
 \Si_i'
&=&
 - \Omega (\Si_i + 1 - 3 w_i)
\,,
\\
 \bar{s}'_i
 &=&
 - 2 (\bar{s}_i+s^\ast_i) \left[ \Si_i  - \sum_k (\bar{s}_k+s^\ast_k) \Si_k \right]
 \,,
\end{eqnarray*}
where $w_i=w_i(\vec s)$.
The corresponding linearized system at $(\vec{\overline{s}},\Si)=(0,0)$ then reads
\begin{subequations}\label{30VII19}
\begin{eqnarray}
 \Si'_i
&=&
 - \Si_i 
 +
 3 \bar{s}_i Y_i(s^\ast_k)
 +
 3 s^\ast_i \sum_j \left. \frac{\partial Y_i}{\partial s_j} \right|_{s^\ast_k} \bar{s}_j
 +
 O(|\Si|^2_\delta) 
 +
 O(|\bar{s}|^2_\delta) 
\,,
\\
 \bar{s}'_i
 &=&
- 2 s^\ast_i \Si_i  
 +
 2 s^\ast_i \sum_k s^\ast_k  \Si_k 
 +
  O(|\bar{s}|_\delta |\Si|_\delta)
 \,.
\end{eqnarray}
\end{subequations}

\section{Energy method}\label{S: proof}
We derive in the following the energy controlling the perturbation away from the fixed point. Note that the $C$ and $\varepsilon$ denote some positive constants that may change from line to line throughout the present paper.

We define the energy by
\begin{equation}
 E_i
 :=
 \alpha
 \bar{s}_i^2
 +
 \beta \bar{s}_i \Si_i
 +
 (\Si_i)^2
 \,,
\end{equation}
where $\alpha > \beta^2 / 4$ and $\beta \in \mathbb{R}$, in order to have $E>0$. 
Now, we pass to our system \eqref{30VII19}. We note that $Y_i(s^\ast_k)=:A_i$ is a constant which can be written as $A_i=1 + a_i$ with $|a_i| = |\varepsilon_i| \ll1$, since $Y_i(s^\ast_k)$ deviates from identity which is the value of it with isotropic distribution function in the background, for small data. Similarly, we set $s^\ast_i = 1/3 + c_i /3$ for $|c_i| = |\varepsilon_i| \ll 1$. On the other hand, we have
\begin{eqnarray*}
 \frac{\partial Y_i}{\partial s_j}
 &=&
  -
   \frac{\int f_0 p_i^2 p_j^2 \left( \sum_k s_k p_k^2 \right)^{-\frac{3}{2}} d^3\tilde{p}}{2 \int f_0  \left( \sum_k s_k p_k^2 \right)^\frac{1}{2} d^3\tilde{p}}
   \nn
   \\
   &&
   -
   \frac{ \int f_0 p_i^2 \left( \sum_k s_k p_k^2 \right)^{-\frac{1}{2}} d^3\tilde{p} \int f_0 p_j^2 \left( \sum_k s_k p_k^2 \right)^{-\frac{1}{2}} d^3\tilde{p}}{ 2 \left(\int f_0  \left( \sum_k s_k p_k^2 \right)^\frac{1}{2} d^3\tilde{p} \right)^2} 
   \,,
\end{eqnarray*}
which is manifestly negative. In particular, if $f_0$ is isotropic, i.e., $f_0(p_i)=f_0(|p|_\delta)$, we obtain by a straightforward calculation
\begin{equation*}
 \left. \frac{\partial Y_i}{\partial s_j} \right|_{s_k=\frac{1}{3}}
 =
  -
 \frac{9}{2} 
 \frac{ \int f_0(|p|_\delta) \, p_i^2 p_j^2  |p|_\delta^{-3} d^3\tilde{p}}{ \int f_0  |p|_\delta d^3\tilde{p}}
 -
 \frac{1}{2}
 =
  \begin{cases}
 -\frac{7}{5}
 \,;
 \quad
 \text{if}
 \quad
 i =j
 \,,
 \\
  -\frac{4}{5}
  \,;
  \quad \text{if}
  \quad i \neq j
  \,.
 \end{cases}
\end{equation*}
Here we used spherical coordinates for $p_i$, i.e., $d^3\tilde{p}= p^2 \sin\theta d p d\theta d\varphi$ with $p:=|p|_\delta$ and used
\begin{eqnarray*}
 \int f_0(|p|_\delta) |p|_\delta d^3 p
 &=&
 4 \pi  \int f_0(|p|_\delta) p^3 dp
 =: 4 \pi I_0
 \,,
 \\
 \int f_0(|p|_\delta) p_i^2 \, |p|^{-1}_\delta \, d^3 p
 &=&
 \frac{4 \pi}{3} I_0
 \,; \quad \forall \, i \in \{1,2,3\}
 \,,
 \\
 \int f_0(|p|_\delta) p_i^2 p_j^2 |p|^{-3}_\delta d^3 p
 &=&
 \begin{cases}
  \frac{4 \pi}{5} I_0
  \,;
  \quad
  \text{if}
  \quad
  i=j
  \,,
  \\
 \frac{4 \pi}{15} I_0
  \,;
  \quad
  \text{if}
  \quad
  i\neq j
  \,.
 \end{cases}
\end{eqnarray*}
This means that for an arbitrary distribution function $f_0$ which deviates slightly from the isotropic one, i.e., $f_0=f_0(|p|_\delta) + \xi \tilde{f}_0$ with $|\xi| \ll 1$, one has
\begin{equation}
 \left. \frac{\partial Y_i}{\partial s_j} \right|_{s_k=s^\ast_k}
 =
  \begin{cases}
 -\frac{7}{5}
 +
 \epsilon_1
 \,;
 \quad
 \text{if}
 \quad
 i =j
 \,,
 \\
  -\frac{4}{5}
  +
   \epsilon_2
  \,;
  \quad \text{if}
  \quad i \neq j
  \,,
 \end{cases}
\end{equation}
where $|\epsilon_1|, |\epsilon_2| \ll 1$. 

To proceed further, we compute
\begin{eqnarray*}
 \frac{d}{d\tau} \Si_i^2
 &=&  
 - 2 \Si_i^2
 +
 6 A_i \bar{s}_i \Si_i
 +
 6
 s^\ast_i \Si_i \sum_j  \left. \frac{\partial Y_i}{\partial s_j} \right|_{s^\ast_k} \bar{s}_j
 +
  O(|\Si|^3_\delta) 
 +
 O(|\Si|_\delta |\bar{s}|^2_\delta)
 \\
 &=&
  - 2 \Si_i^2
 +
 6  \bar{s}_i \Si_i
 +
 6 a_i \bar{s}_i \Si_i
 +
  6
 s^\ast_i \Si_i \sum_j  \left. \frac{\partial Y_i}{\partial s_j} \right|_{s^\ast_k} \bar{s}_j
 +
 O(|\Si|^3_\delta) 
 +
 O(|\Si|_\delta |\bar{s}|^2_\delta)
 \,,
\\ 
  \frac{d}{d\tau} \bar{s}_i^2
  &=& 
  - 4 s^\ast_i \bar{s}_i \Si_i
  +
  4 s^\ast_i \bar{s}_i \sum_k s^\ast_k \Si_k
  +
  O(|\Si|_\delta |\bar{s}|^2_\delta)
 \\ 
 &=&
 - \frac{4}{3}  \bar{s}_i \Si_i
 - 
 \frac{4}{3} c_i \bar{s}_i \Si_i
  +
  4 s^\ast_i \bar{s}_i \sum_k s^\ast_k \Si_k
  +
  O(|\Si|_\delta |\bar{s}|^2_\delta)
  \,,
  \\
  \frac{d}{d\tau} (\bar{s}_i \Si_i)
 &=& 
 - \bar{s}_i \Si_i
 +
 3 A_i \bar{s}_i^2
 -
 2 s^\ast_i \Si_i^2
 +
 2 s^\ast_i \Si_i \sum_k s^\ast_k \Si_k
 +
 3 s^\ast_i \bar{s}_i  \sum_j \left. \frac{\partial Y_i}{\partial s_j} \right|_{s^\ast_k} \bar{s}_j
 +
 O(|\bar{s}|^3_\delta)
 +
 O(|\Si|^2_\delta |\bar{s}|_\delta)
 \\
 &=&
 - \bar{s}_i \Si_i
 +
 3  \bar{s}_i^2
   -
 \frac{2}{3} \Si_i^2
 +
 3 a_i \bar{s}_i^2
 -
 \frac{2}{3} c_i \Si_i^2
 +
 2 s^\ast_i \Si_i \sum_k s^\ast_k \Si_k
 +
 3  s^\ast_i \bar{s}_i  \sum_j \left. \frac{\partial Y_i}{\partial s_j} \right|_{s^\ast_k} \bar{s}_j
 \\
 &&
 +
 O(|\bar{s}|^3_\delta)
 +
 O(|\Si|^2_\delta |\bar{s}|_\delta)
 \,.
\end{eqnarray*}

Then,
\begin{equation*}
 \frac{d}{d\tau} E_i
= 
 3 \beta \bar{s}_i^2
 -
 2 \left( 1 + \frac{1}{3} \beta \right) \Si^2_i
 +
 \left( 6 - \beta - \frac{4}{3} \alpha  \right) \bar{s}_i \Si_i
 +
 \left( \beta \bar{s}_i +  2 \Si_i \right) \sum_j \left. \frac{\partial Y_i}{\partial s_j} \right|_{s^\ast_k} \bar{s}_j
 +
 \mathcal{H}_i(a_i, c_i)
 \,,
\end{equation*}
where
\begin{equation*}
 \mathcal{H}_i(a_i, c_i)
 :=
 3 \beta a_i \bar{s}_i^2
 +
  2  ( 3 a_i - \frac{2}{3} \alpha c_i) \bar{s}_i \Si_i
 -
 \frac{2}{3} \beta c_i \Si_i^2
 +
  \frac{2}{3} (2 \alpha  \bar{s}_i + \beta \Si_i) s^\ast_i  \sum_k c_k \Si_k
  +
 ( 2 \Si_i + \beta \bar{s}_i) c_i \sum_j \left. \frac{\partial Y_i}{\partial s_j} \right|_{s^\ast_k} \bar{s}_j
 \,,
\end{equation*}
which should be taken into account later separately. We now consider the total energy, i.e,
\begin{equation}
 E := E_1 + E_2 + E_3
 \,.
\end{equation}
Using the definitions 
$$
\bar{s}^2 := \sum_i \bar{s}_i^2
\,,
\quad
\vec{\bar{s}} \cdot \vec{\Si}
:=
\sum_i \bar{s}_i \Si_i
\,,
\quad
\mathcal{H}
:=
\sum_i \mathcal{H}_i(a_i, c_i)
\,,
$$
one can easily compute the derivative of the total energy
\begin{eqnarray*}
 \frac{d E}{d\tau}
 &=& 
 3 \beta \bar{s}^2
 -
 2 \left( 1 + \frac{1}{3} \beta \right) F
 +
 \left( 6 - \beta - \frac{4}{3} \alpha  \right) \vec{\bar{s}} \cdot \vec{\Si}
 +
  \sum_{i,j}  \left( \beta \bar{s}_i +  2 \Si_i \right)  \left. \frac{\partial Y_i}{\partial s_j} \right|_{s^\ast_k} \bar{s}_j
 +
 \mathcal{H}
 \\
 &=&
 3 \beta \bar{s}^2
 -
 2 \left( 1 + \frac{1}{3} \beta \right) F
 +
 \left( 6 - \beta - \frac{4}{3} \alpha  \right) \vec{\bar{s}} \cdot \vec{\Si}
 -
 \beta 
  \left[
    \frac{7}{5} \bar{s}^2
   +
   \frac{8}{5}
    \left(
     \bar{s}_1 \bar{s}_2
     +
      \bar{s}_1 \bar{s}_3
      +
       \bar{s}_2 \bar{s}_3
    \right)
  \right]
 \\
 &&
 -
 2
 \left[
  \frac{7}{5}
  \vec{\bar{s}} \cdot \vec{\Si}
  +
  \frac{4}{5}
  \left(
   \Si_1 \left( \bar{s}_2 + \bar{s}_3  \right)
   +
   \Si_2 \left( \bar{s}_1 + \bar{s}_3  \right)
   +
    \Si_3 \left( \bar{s}_1 + \bar{s}_2  \right) 
  \right)
 \right] 
 +
 \mathcal{G}(\epsilon_1, \epsilon_2)
 +
 \mathcal{H}
 \,,
\end{eqnarray*}
where
$$
 \mathcal{G}(\epsilon_1, \epsilon_2)
 :=
 \epsilon_1
  \left(
     \beta  \bar{s}^2
     +
     2 \vec{\bar{s}} \cdot \vec{\Si}
 \right)
 +
 2 \epsilon_2 
 \left[
 \beta
    \left(
     \bar{s}_1 \bar{s}_2
     +
      \bar{s}_1 \bar{s}_3
      +
       \bar{s}_2 \bar{s}_3
    \right)
    +
   \Si_1 \left( \bar{s}_2 + \bar{s}_3  \right)
   +
   \Si_2 \left( \bar{s}_1 + \bar{s}_3  \right)
   +
    \Si_3 \left( \bar{s}_1 + \bar{s}_2  \right) 
  \right]  
  \,.
$$
Using the fact that $\Si_1+\Si_2+\Si_3 = 0$, one can simplify the following term
\begin{equation*}
  \frac{7}{5}
  \vec{\bar{s}} \cdot \vec{\Si}
  +
  \frac{4}{5}
  \left(
   \Si_1 \left( \bar{s}_2 + \bar{s}_3  \right)
   +
   \Si_2 \left( \bar{s}_1 + \bar{s}_3  \right)
   +
    \Si_3 \left( \bar{s}_1 + \bar{s}_2  \right) 
  \right)
  =
  \frac{3}{5}
  \vec{\bar{s}} \cdot \vec{\Si}
  \,,
\end{equation*}
which results in
\begin{equation*}
  \frac{d E}{d\tau}
  =
 \frac{8}{5} \beta  \bar{s}^2
 -
 2 \left( 1 + \frac{1}{3} \beta \right) F
 +
 \left( \frac{24}{5} - \beta - \frac{4}{3} \alpha  \right) \vec{\bar{s}} \cdot \vec{\Si}
 -
   \frac{8}{5}  \beta 
    \left(
     \bar{s}_1 \bar{s}_2
     +
      \bar{s}_1 \bar{s}_3
      +
       \bar{s}_2 \bar{s}_3
    \right)
 +
 \mathcal{G}(\epsilon_1, \epsilon_2)
 +
 \mathcal{H}
 \,.
\end{equation*}
Using the binomial inequality and introducing a decay inducing coefficient $\lambda > 0$ we get
\begin{equation}\label{dE/dtau}
 \frac{d E}{d\tau}
  \leq
  - \lambda E
 +
 q(\vec{\bar s},\vec\Si)
 +
 \mathcal{G}(\epsilon_1, \epsilon_2)
 +
 \mathcal{H}
 \,,
\end{equation}
with the quadratic form
\begin{equation}\label{13VIII19}
 q( \vec{\bar{s}}, \vec{\Si} )
  :=
  \left(
  \lambda \alpha
  +
  \frac{16}{5} \beta
 \right)
 \bar{s}^2
 +
 \left(
  \lambda - 2 - \frac{2}{3} \beta
 \right)
 F
 +
 \left[
  \beta (\lambda -1)
  +
  \frac{24}{5}
  -
  \frac{4}{3} \alpha
 \right]
 \vec{\bar{s}} \cdot \vec{\Si}
 \,.
\end{equation}
We have to choose constants $\alpha$, $\beta$, and $\lambda$ such that the following conditions hold:
\begin{enumerate}[(i)]
 \item  
 $ \alpha > \frac{\beta^2}{4}$, $\beta \in \mathbb{R}$, and $\lambda >0$,
 
 \item $q$ is negative semidefinite.
\end{enumerate}
Then, under these conditions $q$ can be neglected in the inequality \eqref{dE/dtau}. Doing this we arrive at a set of triple numbers $(\alpha, \beta, \lambda)$ which satisfies the two conditions above. To optimize the decay rate we wish to find the triple with maximal value of $\lambda$. It turns out that the optimal value of $\lambda$ is achieved when $q=0$. 
In other words, if $e_1$ and $e_2$ are the two eigenvalues of the quadratic form \eqref{13VIII19}, then the system which corresponds to the conditions (i) and (ii), i.e.,
$$
 \left\{
  \alpha > \frac{\beta^2}{4}
  \,,
  \quad
  \lambda >0
  \,,
  \quad
  e_1 = 0
  \,,
    \quad
  e_2 = 0  
 \right\}
 \,,
$$
is satisfied for the value of $\lambda$ by solving the equation
\begin{equation}\label{1VIII19}
 15 \lambda^3 - 45 \lambda^2 + 142 \lambda -128   = 0
  \,,
\end{equation}
One finds the numerical result $\lambda \approx 1.16426$. Accordingly, $\alpha \approx 3.4455$ and $\beta \approx -1.2536$.

Hence, for this choice of constants we have 
\begin{equation}
 E' \leq - \lambda E + C \varepsilon E
 \,,
\end{equation}
where we estimated $\mathcal{H}$ and $\mathcal{G}$ using suitable constant $C>0$ and  $ 0 <\varepsilon  \ll 1$. This in turn means
\begin{equation}
 \bar{s}_i \leq e^{(- \frac{\lambda}{2} + \varepsilon )  \tau}
 \,,
 \quad
 |\Si_i| \leq e^{(- \frac{\lambda}{2} + \varepsilon ) \tau}
 \,.
\end{equation}

We should express the results in terms of our original time $t$. To this end, we rewrite \eqref{1VIII19.2} as follows
\begin{equation}
 \frac{dH}{dt} 
 =
 - H^2 (2 + \frac{1}{6} F)
 \,, 
\end{equation}
or equivalently
\begin{equation}
 \partial_t (H^{-1})
 =
 2 + \frac{1}{6} F
 \,.
\end{equation}
Since we know that $F \leq C t^{- \kappa}$ for some positive constants $C$ and $\kappa$ which is smaller than one, we have
\begin{equation}
 2 \leq  \partial_t (H^{-1})
 \leq
 2 + C t^{- \kappa}
 \,.
\end{equation}
Integrating this inequality and keeping in mind that $t_0=(2H(t_0))^{-1}$, we arrive at
\begin{equation}
 2t
 \leq
 H^{-1}
 \leq
 2t 
 +
 \frac{C}{1 - \kappa} t^{1 - \kappa}
 \,,
\end{equation}
or equivalently,
\begin{equation}\label{2VIII19}
 H = \frac{1}{2} t^{-1} \left[ 1 + O(t^{-\kappa}) \right]
 \,.
\end{equation}
Now, using $d\tau/dt = H$ and integrating \eqref{2VIII19} we finally find
\begin{equation}
 t^{1/2}
 =
 t_0^{1/2}
 e^{\tau - \tau_0 + \zeta}
 \,,
\end{equation}
where $\zeta := O(\varepsilon (t^{-\kappa}+t_0^{-\kappa}))$ which is a small number.

Therefore, we have
\begin{equation}
  \left| \Si_i \right|
 =
 O(t^{-  \frac{\lambda}{4} + \varepsilon })
 \,,
 \quad
 \left| s_i - s^\ast_i \right|
 =
  O(t^{- \frac{\lambda}{4} + \varepsilon  })
 \,,
 \quad
 \forall i \in \{1,2,3\}
 \,,
\end{equation}
where $\lambda$ is the solution of \eqref{1VIII19}. This ends the proof of the main theorem.

\end{document}